\documentstyle[preprint,prd,aps,epsf]{revtex}
\begin{document}
\draft
\preprint{Imperial/TP/97-98/8}
\title{Applying the linear $\delta$ expansion  to $i\phi^3$}
\author{M. P. Blencowe, H. F. Jones, and A. P. Korte}
\address{The Blackett Laboratory, Imperial College of Science, Technology and 
Medicine,
London~SW7~2BZ, United Kingdom}
\date{\today}
\maketitle
\begin{abstract}
The linear $\delta$ expansion (LDE) is applied to the Hamiltonian
 $H = {1\over 2} (p^2 + m^2 x^2)+ igx^3$, which arises in the study
of Lee--Yang zeros in statistical mechanics.  Despite being non--Hermitian,
this Hamiltonian appears to possess a real, positive spectrum.
In the LDE, as in perturbation theory, the eigenvalues are naturally real, so 
 a proof of this property devolves on the convergence of the expansion.
A proof of convergence of a modified version of the LDE is provided
for the $ix^3$ potential in zero dimensions. The methods developed in zero 
dimensions are then extended to quantum mechanics, where we provide numerical
evidence for convergence.
\end{abstract}


\section{Introduction}

In a recent paper \cite{bender}, Bender and Milton carried out an investigation
of the following non-Hermitian Hamiltonian:
\begin{equation}
H = {1\over 2}(p^2+m^2x^2)+igx^3.
\label{hamiltonian}
\end{equation}
Among other results, they
provided analytical evidence for the remarkable property
that its eigenvalues are real and positive.   
This Hamiltonian and its field theory counterparts arise in the study of the
zeros of the Ising model partition function as a function of complex
magnetic field, the so-called Lee-Yang zeros (see, e.g., Ref.\
\cite{drouffe} and references therein).  
The distribution of zeros in the complex parameter plane of the 
partition function of a given system can  
yield useful information concerning its phase
transitions. For the Ising model, the zeros  lie on the 
imaginary magnetic field axis and, for imaginary field, the effective
theory is an interacting scalar field theory with dominant interaction 
$i\phi^3$ for small fluctuations\cite{bessis}.

In order to give a proper analytical demonstration of the reality of the
eigenvalues of Hamiltonian\ (\ref{hamiltonian}), as well as investigate
the higher dimensional quantum field theory analogues, 
we require an expansion method  which
converges for finite, but otherwise 
arbitrary values of the coupling $g$ and mass $m$. The method must
therefore be necessarily nonperturbative in these parameters. Bender and
Milton employed a variant of their previous $\delta$ expansion method 
\cite{bender2}, in which the interaction $i\phi^3$ is replaced by $(i\phi)^{2+
\delta}$. A Taylor expansion in $\delta$ of the desired quantity 
($N$-point Green function, $n$th. energy eigenvalue, etc.) is obtained and
then $\delta$ is set equal to one. The  terms of the resulting series are
nonperturbative in $m$ and $g$. However, in this method it is extremely
difficult to go beyond first order in $\delta$ and hence test for convergence.           
  
In the present paper we employ an alternative expansion method, 
the so-called linear $\delta$ expansion (LDE).
The LDE has been employed
as a non-perturbative approximation method to study 
problems in, for example, $\phi^4$ theory\cite{0d,1d,stancu}, quantum 
chromodynamics\cite{solov,akeyo}, relativistic nuclear models\cite{pinto},
and electron dynamics in disordered systems\cite{bk}.
The method involves constructing a modified action which involves
the original action of the theory, $S$, and  
a soluble trial action, $S_0$, containing one or more variational parameters
 $\lambda_{i}$:
\begin{equation}
S_{\delta}=S_0 + \delta (S-S_0).
\end{equation}
The Green function of interest is evaluated as a power series in
 $\delta$, which is then set equal to 1.
 For the success of the method the trial action
 $S_0$ needs to be simple, so that one can perform high order calculations,
but also as close to the true action as possible, so that expanding around 
 $S_0$ is a reasonable procedure with a good chance of converging.
Although $S_{\delta =1}=S$ is  independent of the ${\lambda_i}$, there is 
nonetheless 
a residual dependence on ${\lambda_i}$ in the truncated series evaluated
at $\delta=1$, and it is therefore necessary to choose these parameters
according to some well--defined criterion. Perhaps the most commonly
used criterion is  the principle of minimal sensitivity (PMS), according to which
the ${\lambda_i}$ are chosen to be stationary points of the truncated
expansion, where the dependence is minimal. Whatever criterion
is adopted, it is to be applied at each order $N$ in the expansion, so that 
the ${\lambda_i}$ become $N$--dependent.
This feature of the method is crucial for the convergence of the LDE,
which can be characterized as an order--dependent split between the
bare and the interaction terms in the action. In many cases where the fixed
split of conventional perturbation theory leads to a divergent series, the LDE
can be proved to lead to a convergent sequence of approximants. 

In this paper we apply different variants of the LDE to the $ix^3$ potential in 
zero dimensions,
where we give a proof of convergence of the expansion for the analogue
of the vacuum persistence amplitude $Z$. We then go on to consider the 
one--dimensional problem and provide numerical
evidence of convergence of the LDE for the finite-temperature partition function
and the ground-state energy.

In Sec.\ II, which deals with the zero-dimensional problem, we first
describe how the conventional LDE converges, but to the wrong answer.   
One possible resolution is to split the integral up for positive and negative
$x$ and apply the PMS separately to each integral\cite{bender3}.
Numerically, this gives a sequence of approximants
converging to the correct answer, and we provide a proof of
convergence of this procedure. This LDE variant can be generalized
to the path integral expression of the 
1-D finite-temperature partition function. 
However, difficulties are encountered when applied
to the ground-state energy, owing to the non-analyticity of the splitting 
procedure.
We resolve this problem by using a modified $\delta$ expansion which 
involves a shift parameter.  This retains the essential features of the 
integration splitting procedure, but 
has the advantage of being analytic, thus making the
calculations straightforward to carry out and allowing the possibility
of  generalization to higher dimensions. In the last part of Sec.\ II,
a numerical study of the shift method is carried out, and a proof of
convergence provided using saddle-point techniques.

In Sec.\ III, where we deal with the quantum mechanical problem,
we show that the conventional LDE again fails. Numerical evidence of 
convergence is given for the path integral expression of the  partition function
(using both the splitting and shift techniques) and especially for the ground state 
energy using the shift technique.

In the conclusion we outline further directions,
including  proving convergence for the $1$-D problem and generalizing the
shift method to the higher dimensional field theory analogues.


\section{Zero dimensions}

The zero--dimensional analogue of the finite-temperature partition function in 
quantum mechanics, or the vacuum persistence amplitude in field theory, is the
ordinary integral
\begin{equation}
Z=\int_{-\infty}^{\infty} dx e^{-m^2x^2+igx^3}.
\label{Z0exact}
\end{equation}
For simplicity we will take $m=0$. In that case $Z\propto g^{-1/3}$ and
we can set $g=1$ without loss of generality. In spite of the absence of
the convergence factor the integral is still well--defined, and can be
calculated by splitting up the integration range into $x<0$ and
$x>0$, and then rotating the contour by $\pm\pi/6$. The exact result, 
obtained in this way, is $Z=\Gamma(1/3)/\sqrt{3}= 1.54668588415598$, to
15 significant figures. 
\subsection{Na\"{\i}ve application of the LDE}

In evaluating $Z$ using the LDE, the standard approach is to modify the 
exponent simply by adding and subtracting a quadratic term, to give
\begin{equation}
\label{naive}
Z(\delta)=\int_{-\infty}^{\infty} dx e^{-\lambda x^2+\delta (\lambda x^2+ix^3)}.
\end{equation}
The procedure is then to expand $Z_\delta$ to order $\delta^N$, set $\delta=1$ and
then choose $\lambda=\lambda_N$ by some criterion or other. In the absence of any
additional information the most reasonable criterion is the principle of
minimal sensitivity (PMS), namely to choose $\lambda_N$ as a stationary point
of $Z_N$.

The truncated series is $\sum_{n=0}^N c_n$, with
\begin{equation}
c_n=\frac{1}{\sqrt{\lambda}} \sum_{r=0}^{[n/2]}
\frac{\Gamma(n+r+1/2)}{(2r)!(n-2r)!}\left(-\frac{1}{\lambda^3}\right)^r.
\label{zn}
\end{equation}
Using (\ref{zn}), $Z_N$ can be calculated to high order. However, the residual
dependence on $\lambda$ turns out to be radically different from the $x^4$ case
discussed in \cite{0d}, where, for odd $N$ there was a single maximum, and
the value of $Z$ at that maximum steadily approached the exact value as $N$ 
increased. As shown in Fig.~1, where $Z_N(\lambda)$ is plotted against $\lambda$ 
for $N=30$,
the situation here is that there are extremely
violent oscillations for small $\lambda$, which gradually decrease in amplitude as
$\lambda$ increases, with a final, very broad maximum lying above the
exact value of $Z$. 

One's natural choice of a PMS point for $\lambda$ would be the
position of this last maximum: the residual dependence of $Z_N$ on $\lambda$ around
this point is much less than anywhere else on the graph. However, it turns
out that the sequence $Z_N(\lambda_N)$, with $\lambda_N$ so chosen, indeed 
converges, 
but to the wrong answer! The same is true of the previous minimum, and indeed
none of the stationary points converges to the correct answer. This is a
stark warning, in this admittedly somewhat pathological case, of the
shortcomings of the PMS. It is worth mentioning that the scaling behaviour
with $N$ of these $\lambda_N$ is anomalous: for large $N$ they grow like $N$,
rather than the $N^{1/3}$ which would be expected from a saddle-point analysis
and which indeed is obtained in the variants of the LDE discussed in the
following two subsections. In an estimation of the error $Z-Z_N$ on the lines
of the double saddle--point procedure of Ref.~\cite{1d}, the interaction term
$ix^3$ would then be sub--dominant, a clear signal that one is on the wrong
track.

\subsection{The splitting procedure}
A possible resolution of this problem is to split up the integration range into
positive and negative $x$. This gives
$Z_N(\lambda)=Z_N^{+}(\lambda) + Z_N^{-}(\lambda)$, where
\begin{equation}
Z_N^{+}(\lambda)=\int_0^{\infty} dx e^{-\lambda x^2} \{e^{\lambda x^2+ix^3}\}_N,
\end{equation}
and similarly for $Z_N^{-}$, where $\{f(z)\}_N$ denotes the Taylor
expansion of $f$ to order $z^N$.
If $\lambda$ is taken as $\lambda =r e^{-i\pi /3}$, where $r$ is real, and the 
integration contour rotated by $x\rightarrow x\,e^{i\pi/6}$, then
$Z_N^{+}(\lambda)=e^{i\pi /6}{\cal Z}_N(r)$ where
\begin{equation}
{\cal Z}_N(r)=\int_0^{\infty} dx e^{-rx^2}\{e^{rx^2-x^3}\}_N.
\label{znplus}
\end{equation}
For $Z_N^{-}$ we need to take $\lambda =r e^{i\pi /3}$ and rotate the contour
in the opposite direction, to obtain $Z_N^{-}=e^{-i\pi /6}{\cal Z}_N(r)$.

The analysis of ${\cal Z}_N$ is very similar to the $x^4$ case discussed in
Ref.~\cite{0d}. For odd $N$ there is a single maximum in $r$, which steadily
converges to the exact result.  For $N=11$ the situation is depicted in Fig.~2,
where we have multiplied ${\cal Z}_N$ by the appropriate factor of $\surd 3$.
The contrast with Fig.~1 could hardly be more striking. The crucial difference
between this and the na\"{\i}ve application of the LDE is that by splitting the
integral we are in fact using two different (complex conjugate) values of
$\lambda_N$ for the two integrals rather than the same value for both.

The proof of convergence of this procedure is similar to that for
the $x^4$ case, though different in detail. The remainder
 $R_N$ is essentially given by the integral for the $N$th. coefficient
in the expansion of ${\cal Z}$, namely
\begin{equation}
c_N={1 \over N!}\int_0^\infty dx\; e^{-rx^2}(rx^2-x^3)^N,
\end{equation}
which can be estimated by saddle--point methods for large $N$. In this case
the PMS value of $r_N$ scales as $N^{1/3}$. Writing $r_N=\alpha N^{1/3}$ and
 $x=zN^{1/3}$ and using Stirling's approximation for $N!$, the leading
behaviour of the integral is
\begin{equation}
c_N=\int_0^\infty dz\, e^{N\varphi},
\end{equation}
where
\begin{equation}
\varphi = -\alpha z^2 +\ln(\alpha z^2 -z^3)+1.
\end{equation}
The saddle--point equation $d\varphi/dz=0$ is
\begin{equation}
2\alpha z^3 - 2\alpha^2 z^2 -3z +2\alpha = 0,
\end{equation}
which has two positive roots $z_\pm$. What is required for convergence
is that both ${\rm Re}\,\varphi(z_{+})$ and ${\rm Re}\,\varphi(z_{-})$ be negative, 
and then the rate of convergence is governed by the larger of the two. 
They become equal at the PMS point, when $\alpha$ is such that 
 ${\rm Re}\,\varphi(z_{+})={\rm Re}\,\varphi(z_{-})$. 
Numerically the solution of this equation is $\alpha=1.0272$, 
when ${\rm Re}\,\varphi(z_\pm)=-1.2398$. 
Thus the sequence of approximants converges like $\exp(-1.2398 N)$.

\subsection{The shift method}

If we were only concerned with zero dimensions this would be an
entirely satisfactory resolution of the problem. However, in higher
dimensions one is dealing with a {\it functional} integral, and the
generalization of such a splitting procedure is fraught with
difficulties, even though some progress can be made along these lines,
as we show in Sec.~IIIB. Fortunately, consideration of the quantum
mechanical problem suggests another solution, which is immediately
generalizable to higher dimensions, namely to incorporate a linear
term, or shift, into the $\delta$--modified action.

The motivation for introducing such a term is discussed briefly
in Sec.~IIIC. In the context of the zero--dimensional model it
is worth noting that one of the features of the successful stratagem
of splitting the integration range and treating each half separately
is that terms odd in $x$ survive the integration, whereas they cancel
in the na\"{\i}ve application of the $\delta$--expansion. 
Introducing a linear term into the $\delta$--modified action also 
avoids such a cancellation, but in a simple algebraic way.

Thus, in zero dimensions the relevant modification of Eq.~(\ref{naive})
is
\begin{equation}
Z(\delta)=\int_{-\infty}^{\infty} dx e^{-\lambda x^2+\delta [\lambda x^2+
ig(x+ia)^3]},
\end{equation}
where we have introduced a shift $ia$, which will indeed turn out
to be pure imaginary. Notice that when $\delta=1$ this does reduce
to the original integration, as the shift is then immaterial.

Truncating the expansion at order $\delta^N$ and setting $\delta=1$
we now have the series $\sum_{n=0}^N c_n$, where
\begin{equation}
c_n=\sum_{r=0}^{n} \sum_{s=0}^{[3r/2]}
{3r\choose 2s}
\frac{\Gamma(n-r+s+1/2)}{r!(n-r)!\lambda^{s+1/2}}(-1)^s g^r
a^{3r-2s}.
\end{equation}

One can now search for double PMS points in the two parameters $a$, $\lambda$.
For $N>2$ there are several such points. For example, at order 16 there
are ten solutions, with $\lambda$ lying in the range 4--6 and $a$ close to 1.
There is no very convincing criterion for choosing between these multiple
PMS solutions, but they all agree with the exact answer up to the 7th.
decimal place.

However, the PMS was only a means to an end, namely to obtain a sequence
of approximants converging to the exact answer, and in the present case
we can instead adopt this latter requirement directly as a criterion for
choosing $a$ and $\lambda$. This can be done via a saddle--point analysis of
the error $R_N$.

As in the previous subsection, the error is essentially given by the
expression for the $N$th. coefficient $c_N$, which in the present case
reads
\begin{equation}
c_N={1\over N!}\int_{-\infty}^{\infty} dx\,e^{-\lambda x^2}
\left[ \lambda x^2 + ig(x+ia)^3\right]^N.
\end{equation}
The appropriate scaling for a saddle--point approximation in which
both parameters play a non--trivial r\^ole is $\lambda=\alpha N^{1/3}$, 
 $a=\beta N^{1/3}$. Then the large--$N$ behaviour of $c_N$ is
\begin{equation}\label{cnsh}
c_N = \int_{-\infty}^\infty dz\,e^{N\varphi},
\end{equation}
where now (again setting $g=1$)
\begin{equation}
\varphi=-\alpha z^2 +\ln\left[\alpha z^2 +i(z+i\beta)^3\right]+1.
\end{equation}
The saddle--point equation in this case is a quartic:
\begin{equation}
2\alpha z^4 +2i\alpha(3\beta-\alpha)z^3-3(2\beta^2\alpha+1)z^2-
2i(\beta^3\alpha+3\beta-\alpha)z+3\beta^2=0,
\end{equation}
with four complex roots.

The location of these roots is shown in Fig.~\ref{paths} for typical
values of $\alpha$ and $\beta$, when they occur as two pairs of the form
 $\pm u+iv$. Also shown in that diagram are the paths of
stationary phase. It is necessary to know the geometry of these paths
in order to determine which saddle points are encountered, and in what 
direction, when the contour is distorted from its original location along
the real axis from $-\infty$ to $\infty$. In fact the required path 
goes through all four saddle points $A$, $B$, $C$, $D$, in each case
at a maximum of the integrand because of the intervening ``sinks'' $X$, 
$Y$, $Z$ where the argument of the logarithm vanishes.

The values of $\varphi$ at the mirror points $A$ and $B$ are the complex 
conjugates of those at $D$ and $C$ respectively. The criterion for
convergence therefore reduces to ensuring that both ${\rm Re}\,\varphi(z_C)$
and ${\rm Re}\,\varphi(z_D)$ are negative. Optimal convergence occurs when
the two are equal.

Thus in the parameter space of $\alpha$ and $\beta$ we solve the equation
  ${\rm Re}\,\varphi(z_C) = {\rm Re}\,\varphi(z_D)$ and then look for points
where this common value is negative. We display the results as a contour plot 
in Fig.~\ref{cntr}, where the outer contour corresponds to
 ${\rm Re}\,\varphi(z_C)=0$ and the inner one to ${\rm Re}\,\varphi(z_C)=-1$.
Any point within the outer contour will give convergence, while points
within the inner contour will give convergence with an error smaller
than $e^{-N}$. Notice that there is quite a wide range of
acceptable values of $\beta \equiv aN^{-1/3}$, but of course this does not
include $\beta=0$, which would correspond to the na\"{\i}ve LDE with no
shift term.

In Fig.~\ref{cgce} we demonstrate the nature of the numerical
convergence up to $N=30$ for the optimal values $\alpha=1.79055$,
 $\beta=0.38378$. The abolute value of the error is shown, on a logarithmic
scale. Although this decrease lies within the predicted envelope, the error
is periodically significantly smaller than the general trend. A similar
pattern of convergence has been seen\cite{hk} in the application of 
variational
perturbation theory to the calculation of the strong--coupling 
coefficients of the quartic oscillator.

In conclusion, the shift method described in this subsection involves
a simple algebraic modification of the original integrand and an
unambiguous choice of the two variational parameters according to the
requirement of convergence. The method may be straightforwardly 
generalized to higher dimensions, where the calculations involved in
the $\delta$ expansion are not significantly more complicated than in 
ordinary perturbation theory. In the case of quantum mechanics the
analytic nature of the $\delta$--modified Hamiltonian allows one to go to
very high orders, as we show in subsection IIIC.

\section{One dimension}

We now return to the quantum mechanical problem armed with the insights
gained from the zero--dimensional analogue.

The initial evidence for the reality and positivity of the spectrum comes
from a numerical analysis using the matrix method, whereby the Hamiltonian
is rewritten in terms of the raising and lowering operators of the bare
Hamiltonian and then regarded as a matrix $H_{mn}$ in the infinite--dimensional 
space labelled by the occupation numbers $n$. If this matrix is truncated
at some finite order $N$ the resulting spectrum of eigenvalues can easily
be calculated numerically, up to $N$ of the order of 100. The pattern which
emerges from such a calculation is that as $N$ increases more and more of 
the lower eigenvalues emerge as real numbers from the amalgamation of complex
conjugate pairs. After emergence they each tend to a definite positive limit.
The ground--state energy is always present from $N=3$ onwards, and for $m=g=1$
is stable at $E_0=0.797342612$ to the 9th. decimal place beyond $N=40$.

Since the $E_i$ appear to be real, so also will be the finite--temperature
partition function $Z(\beta)=\sum_i \exp(-\beta E_i)$, which can be calculated 
to the desired accuracy by including a sufficient number of energy levels in 
the sum.

To our knowledge there is no proof of the convergence of the matrix method, and
our ultimate goal will be to provide such a proof for a suitable generalization
of the LDE. In this section we discuss in turn the three variants of the 
$\delta$ expansion which were used in the zero--dimensional analogue.

\subsection{Na\"{\i}ve application of the LDE}

The standard $\delta$--modification of Eq.~(\ref{hamiltonian}) amounts
basically to expanding around a SHO of a different, order--dependent
frequency according to
\begin{equation}
H={1\over 2}\left[p^2+(m^2+2\lambda)x^2\right]+\delta (igx^3-\lambda x^2).
\end{equation} 
The new effective interaction is $igx^3-\lambda x^2$, a polynomial.
For a polynomial $V(x)$ one has the possibility of expanding to very
high orders via the use of recursion relations rather than the much
more cumbersome Rayleigh--Schr\"odinger perturbation theory.

These recursion relations are derived by first factoring off the asymptotic
behaviour of the wave--function according to $\psi = e^{-{1\over 2} y^2}\chi$,
where $y=x\surd\omega$ and $\omega=(m^2+2\lambda)^{1\over 2}$. Then $\chi$ and the 
ground--state energy are both expanded as power series in $\delta$, with 
$\chi=1+\sum_{n=1}^\infty \delta^n\chi_n$ and
$E={1\over 2}\omega(1+\sum_{n=1}^\infty \delta^n\epsilon_n$). The crucial 
observation is
that with a polynomial interaction of order 3, the functions $\chi_n$
are polynomials of order $3n$, so that we can write $\chi_n=\sum_{p=1}^{3n}
A_{n,p}(iy)^p$. Substituting these expressions into the Schr\"odinger
equation gives coupled recursion relations for the $\epsilon_n$ and $A_{n,p}$, 
with $\epsilon_n=2A_{n,2}$.

Using MACSYMA we have evaluated $E$ up to order $N=20$.
The results are similar to those obtained for $Z$ in zero dimensions,
namely there are very large oscillations for small $\lambda$, while for
larger $\lambda$ there are a number of PMS points, with $\lambda$ scaling like
$N$, where the values of $E$ tend to a constant different from the
exact value.

\subsection{The splitting procedure}

In Sec.\ IIB, we saw how applying LDE separately to $Z^+$ and $Z^-$ gave a
sequence which converged to the correct answer. A natural generalization of
the splitting procedure to the path integral formulation of the
quantum-mechanical partition function is given by $Z_{\delta}=Z_{\delta}^{+} 
+ Z_{\delta}^{-}$, where
\begin{equation}
Z^{\pm}_{\delta}=\int [dx]\  \theta \left(\pm
\int^{\beta /2}_{-\beta /2} x dt\right)
\exp\left( -\int^{\beta /2}_{-\beta /2} dt \left\{
{1\over 2}\left[\dot{x}^2+(m^2+2\lambda_{\pm})x^2\right]+\delta (igx^3-
\lambda_{\pm}x^2)\right\}\right).
\label{zqmpm}
\end{equation}
This expression is rather formal as it stands, since it is not possible
to evaluate the partition function  directly in the presence of the step
function. A possible resolution \cite{bk} is to use the 
integral 
representation of the step function:    
\begin{equation}
\theta (z)={1\over 2\pi i}\int _{-\infty}^{\infty} dq \frac{e^{iqz}}{q-i\epsilon}
\label{intstep}
\end{equation}
which leads to
\begin{equation}
Z^{\pm}_{\delta}={1\over 2\pi i}\int _{-\infty}^{\infty} dq 
\frac{Z^{\pm}_{\delta}(q)}
{q-i\epsilon},
\label{properzpm}
\end{equation}
where
\begin{equation}
Z^{\pm}_{\delta}(q)=
\int [dx]\
\exp\left( -\int^{\beta /2}_{-\beta /2} dt \left\{
{1\over 2}\left[\dot{x}^2+(m^2+2\lambda_{\pm})x^2\right]\mp iqx+\delta (igx^3-
\lambda_{\pm}x^2)\right\}\right).
\label{zpmq}
\end{equation}
We have checked that  the zero--dimensional analogue of (\ref{properzpm}) gives the 
same result as splitting up the integral directly.    

To first order in $\delta$, Eq.~(\ref{properzpm}) gives (cf. \cite{zinn})
\begin{equation}
Z^{\pm}={1\over 2}\bar{Z}^{\pm}\left\{1+\beta\left[\lambda_{\pm}\Delta_\pm \mp
{2ig\over \surd(2\pi)\omega_\pm}(3\Delta_\pm-1/\omega_{\pm}^2)\right]\right\}
\label{zqm}
\end{equation}
where 
$\omega_{\pm}^2=m^2+2\lambda_{\pm}$,
 $\Delta_{\pm}=\coth({1\over 2}\omega_{\pm}\beta)/2\omega_{\pm}$, 
and $\bar{Z}^{\pm}=1/[2\sinh ({1\over 2}\omega_{\pm} \beta) ]$
is the partition function for the simple harmonic oscillator.

For each of $Z^\pm$ there is a single complex PMS point, with 
$\lambda_{-}=\lambda_{+}^*$ and $Z^{-}=(Z^{+})^*$. A similar
calculation can be carried out to order $\delta^2$, where in each
case we find two complex PMS points.
In Fig.\ \ref{zsplit}
we plot the order $\delta$ and $\delta^2$ approximations to 
the finite-temperature partition function as a function of the
coupling $g$ for $\beta =2$ and $m^2=1$, and compare them 
with the partition function calculated via $Z=\sum_i e^{-\beta E_i}$, 
with the energy eigenvalues\footnote{For $\beta=2$ sufficient accuracy
was attained by including at most 10 energy levels in the sum.}
obtained by the matrix method. Both possible solutions
at $O(\delta^2)$ are given, and it will be seen that they do not differ 
appreciably from each other. The graph provides some evidence
for the convergence of this variant of the LDE method, but it is 
obviously not conclusive. However, it is difficult to go to significantly
higher orders in this approach.

\subsection{The shift method}
While we have seen that the splitting method can be implemented 
in the calculation of $Z$ via 
the integral representation of the theta function, calculations to 
higher orders become progressively more
complicated. Moreover, there is no obvious way to implement the method
for the calculation of the energy levels themselves. The shift method,
in contrast, can be applied to both $Z$ and the energy levels, and in
the latter case recursion relations allow one easily to go to high
orders in $\delta$.

In the context of quantum mechanics, the motivation for introducing a
shift arises from a modification of the variational approach for the
calculation of the ground--state energy using a trial Gaussian--like
wavefunction 
\begin{equation}\label{trialwf}
\psi=N[\theta (x) e^{-\lambda_1 x^2}  + 
\theta (-x) e^{-\lambda_2 x^2}]  
\end{equation}
which distinguishes between $x<0$
and $x>0$. Because of the symmetry of the Hamiltonian under
$x\rightarrow -x$ combined with complex conjugation it is clear
that $\lambda_1=\lambda_2^*$.

In the approach of \cite{stevenson} a delta--modified Hamiltonian
is constructed which coincides with the Gaussian variational 
approximation to lowest order in Rayleigh--Schr\"{o}dinger perturbation
theory, but then allows one systematically to improve on the variational
approximation. The difficulty with this approach in the present
problem is the non-analyticity of the Hamiltonian, which considerably 
complicates the calculation of the higher order terms.

The shift method can be regarded as a modification of this approach 
which retains its essential features but involves an analytic
wave--function and Hamiltonian. Thus with $\lambda_1=b+ia$ the
wave--function of Eq.~(\ref{trialwf}) can be written as
 $\psi=Ne^{-(bx^2+iax|x|)}$, with the non--analyticity residing
in the $|x|$ factor in the imaginary part of the exponent, which
is, as a consequence, odd in $x$. The simplest alternative construction
which retains this feature, while at the same time being analytic, is
 $\psi \sim e^{-(bx^2+iax)}$. Thus, a (pure imaginary) shift parameter 
$ia$ has been introduced. A similar shift procedure is in fact described in
\cite{stevenson}, and a method given to $\delta$--modify a Hamiltonian 
incorporating a shift. Following this method we obtain the 
Hamiltonian
\begin{equation}
\label{Hshift}
H_\delta={1\over 2} p^2+\lambda x^2-{1\over 2} m^2 a^2 +ga^3
+\delta \left[{1\over 2} m^2(x+ia)^2-\lambda x^2 +{1\over 2} m^2 a^2 
-ga^3+ig(x+ia)^3\right].
\end{equation}

The corresponding Lagrangian can be used for the calculation of the
finite--temperature partition function
as a functional integral rather than as $\sum_i e^{-\beta E_i}$. The
first--order approximation to $Z$ obtained in this way \cite{zinn} is
\begin{equation}
Z=\bar{Z}e^{\beta({1\over 2} m^2a^2-ga^3)}\left[1+\beta \Delta_0\left(3ga+\lambda-
{1\over 2} m^2\right)\right],
\end{equation}
where $\Delta_0=\coth (\sqrt{2\lambda} \beta/2)/2\sqrt{2\lambda}$
and $\bar{Z} = 1/[2\sinh(\sqrt{2\lambda}\beta/2]$.
We have also calculated the second--order approximation.
In Fig.\ \ref{zshift} we plot both the order $\delta$ and $\delta^2$
approximations to $Z$ as a function of the coupling $g$, where we 
have again taken $m^2=1$ and  $\beta=2$. On the scale of the diagram
the $O(\delta^2)$ approximation is almost indistinguishable from
that obtained from the matrix method, which provides  strong
evidence for the rapid convergence of this variant of the LDE 
for the partition function.

However, the most convincing numerical evidence for the numerical
convergence of the shift method comes from a calculation of the
ground state energy. This can be done by generating recursion relations
for Eq.~(\ref{Hshift}) in a similar manner to that described in
Sec.~IIIA.
 
We have applied the recursion relation method to eighth 
order in $\delta$, taking $m=1$ for definiteness. In  Fig.\ \ref{Eshift} 
we show the results of the $O(\delta)$ and $O(\delta^2)$ calculations for a
range of $g$, while in Table 1 we illustrate the nature of the numerical
convergence up to $O(\delta^8)$ for $g=1$. In the latter case we
consistently chose the largest
PMS value of $\lambda$. The numerical evidence is strong
and encourages us to seek a proof of convergence for the 
one--dimensional case.

\begin{center}
\tabcolsep=25pt
\begin{tabular}[b]{c|lll}
$N$ & $\lambda$ & $a$ & $E_0$\\\hline
1 & 1.817 & 0.3837 & 0.7785\\
2 & 2.169 & 0.4445 & 0.7957\\
3 & 2.301 & 0.4608 & 0.7967\\
4 & 2.426 & 0.4793 & 0.79720\\
5 & 2.525 & 0.4930 & 0.79729\\
6 & 2.6107 & 0.5040 & 0.79732\\
7 & 2.6865 & 0.5135 & 0.797334\\
8 & 2.75475 & 0.52145 & 0.797339
\end{tabular}
\end{center}
\begin{center}
Table 1:\quad Results of the shift method for $E_0$ (to be compared with 
0.797342612\dots)
\end{center}
\section{Conclusion}
We have applied the LDE approximation method to the $ix^3$ potential in 
zero and in one dimension. A proof of convergence was given for the 
massless case 
in zero dimensions. The methods employed in zero dimensions were then 
extended 
to one dimension where numerical evidence of convergence for the
partition 
function and the ground state energy was obtained. This evidence
was particularly strong for the ground--state energy within
the shift method.

The next step is to extend the proof of convergence from zero 
to one dimension. For $Z$ one would need to apply techniques 
similar to those used 
in \cite{1d}, while for the energy levels an extension of the
methods of Guida et al. \cite{Guida} would be necessary.

In higher dimensions one has to tackle the problems of regularization
and renormalization within the LDE. A possible way of doing this has
been outlined by the last--quoted authors in the context of
the calculation of the critical exponents of $\phi_3^4$ 
using the order--dependent mapping method, which is equivalent to
the conventional LDE without a shift. It would be interesting to generalize 
their methods to trial actions including a shift, and to apply
these techniques to the study of Lee--Yang zeros in statistical mechanics.

\section*{Acknowledgements}

We would like to thank Prof. C.~M.~Bender for bringing this problem to
our attention and for assistance during the initial stages of the project.
M.~B. acknowledges financial support from EPSRC.



\begin{figure}
\caption{$Z_N$ for $N=30$ as a function of $\lambda$ in the na\"{\i}ve LDE in
zero dimensions. The dotted line shows the exact answer.}
\label{figa}
\end{figure}

\begin{figure}
\caption{$Z_N$ for $N=11$ as a function of $r$ using the splitting method in
zero dimensions. The dotted line shows the exact answer.}
\label{figb}
\end{figure}

\begin{figure}
\caption{The location of the saddle points and sinks in the integrand
of Eq.~(\protect\ref{cnsh}). The required stationary--phase path, passing through
all four saddle points, is $A\,X\,B\,Y\,C\,Z\,D$.}
\label{paths}
\end{figure}

\begin{figure}
\caption{ Contours of ${\rm Re}\,\varphi$ in the parameter space of
$\alpha$ and $\beta$. Convergence with the shift method is assured for
all points within the outer contour. For points within the inner
contour the convergence is faster than $e^{-N}$}
\label{cntr}
\end{figure}

\begin{figure}
\caption{Convergence of the LDE for the shift method in zero dimensions
for the optimal values of $\alpha$ and $\beta$. The abolute value of the error
is plotted on a logarithmic scale against $N$.}
\label{cgce}
\end{figure}

\begin{figure}
\caption{The finite temperature partition function versus $g$, with fixed
$m^2=1$ and $\beta =2$. The solid line is the numerical matrix method
approximation, the dashed line the order $\delta$ approximation 
and the dotted lines the order $\delta^2$ approximation using the splitting 
procedure.}
\label{zsplit}
\end{figure}

\begin{figure}
\caption{The finite temperature partition function versus $g$, with fixed
$m^2=1$ and $\beta =2$. The solid line is the numerical matrix method
approximation, the dashed line the order $\delta$ approximation 
and the dotted line the order $\delta^2$ approximation using the shift 
procedure.}
\label{zshift}
\end{figure}

\begin{figure}
\caption{The ground state energy versus $g$, with fixed
$m^2=1$. The solid line is the numerical matrix 
method approximation, the dashed line the order $\delta$ approximation 
and the two dotted lines the order $\delta^2$ approximation with two
different PMS values for $\lambda$, using the shift method.}
\label{Eshift}
\end{figure}
\setcounter{figure}{0}
\begin{figure}
\epsfxsize=6in
\epsffile{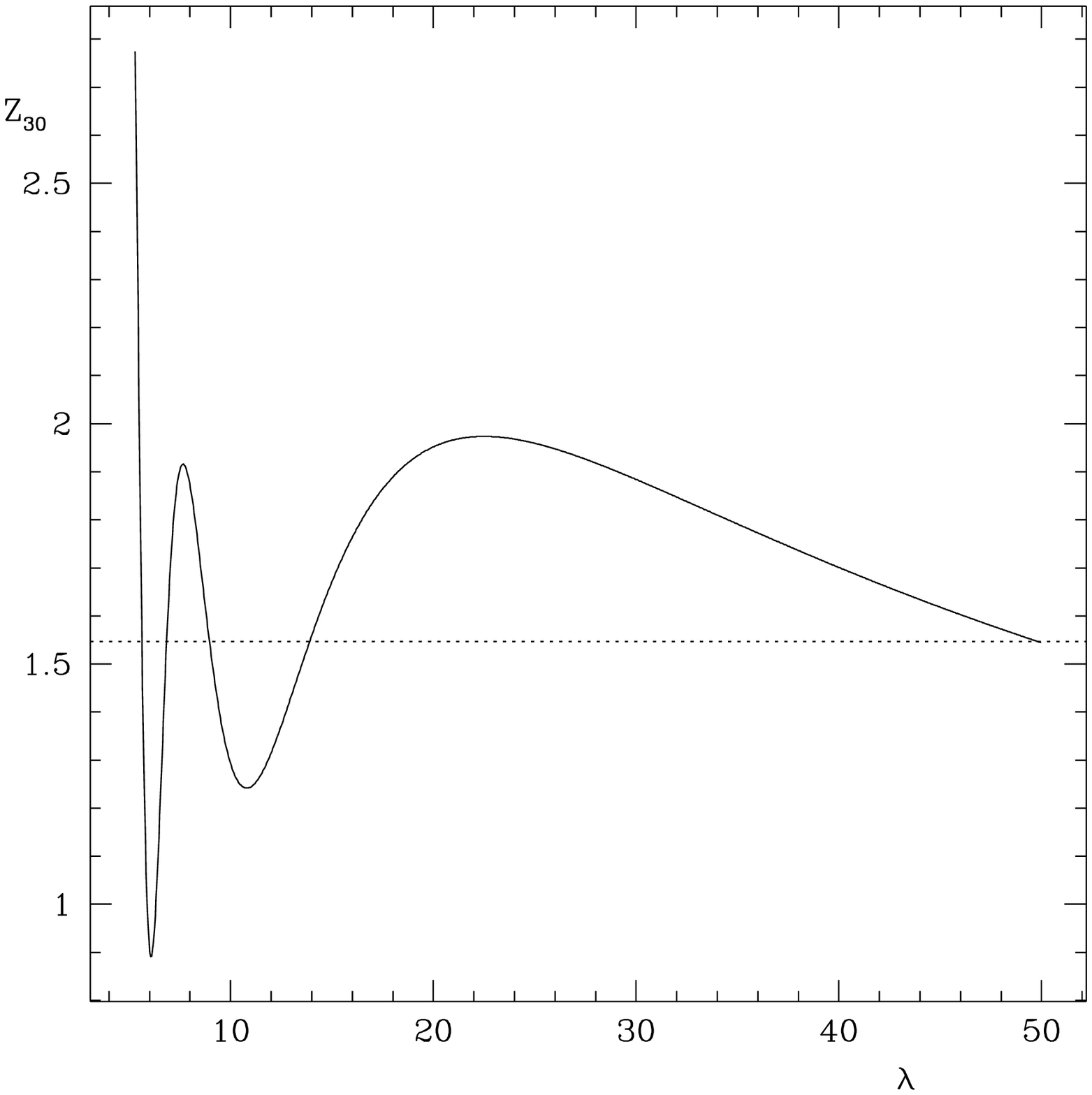}
\caption{}
\end{figure}

\begin{figure}
\epsfxsize=6in
\epsffile{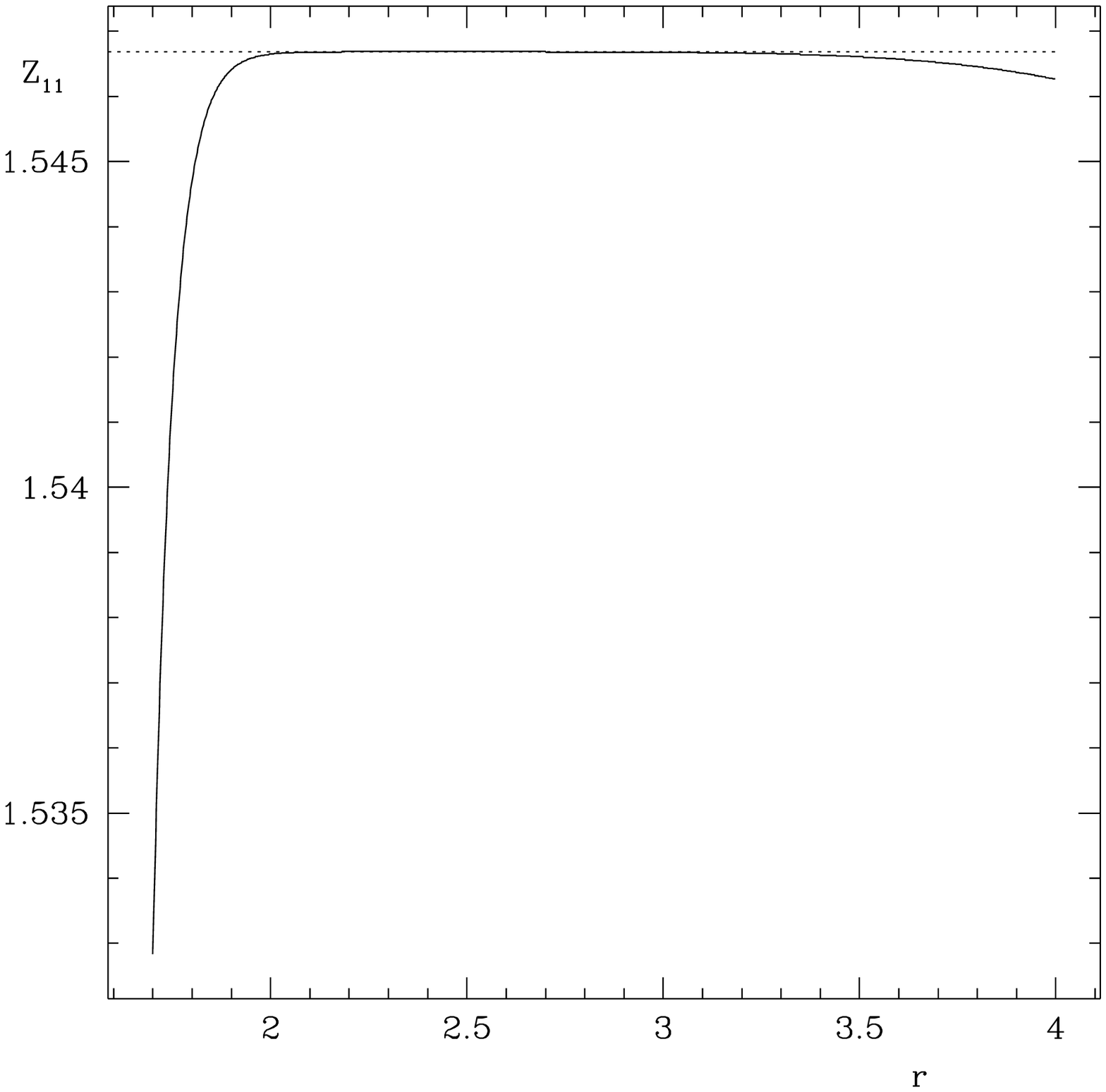}
\caption{}
\end{figure}

\begin{figure}
\epsfxsize=6in 
\epsffile{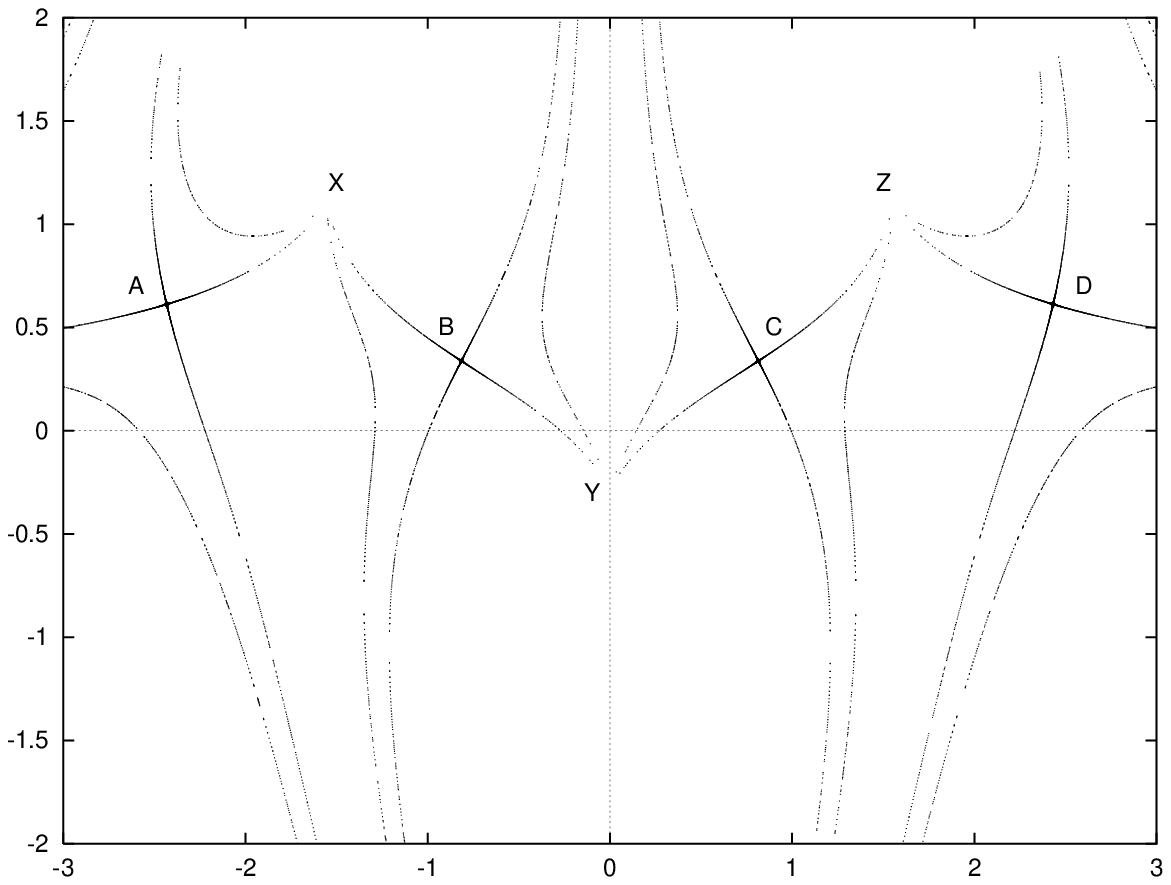}
\caption{}
\end{figure}

\begin{figure}
\epsfxsize=6in
\epsffile{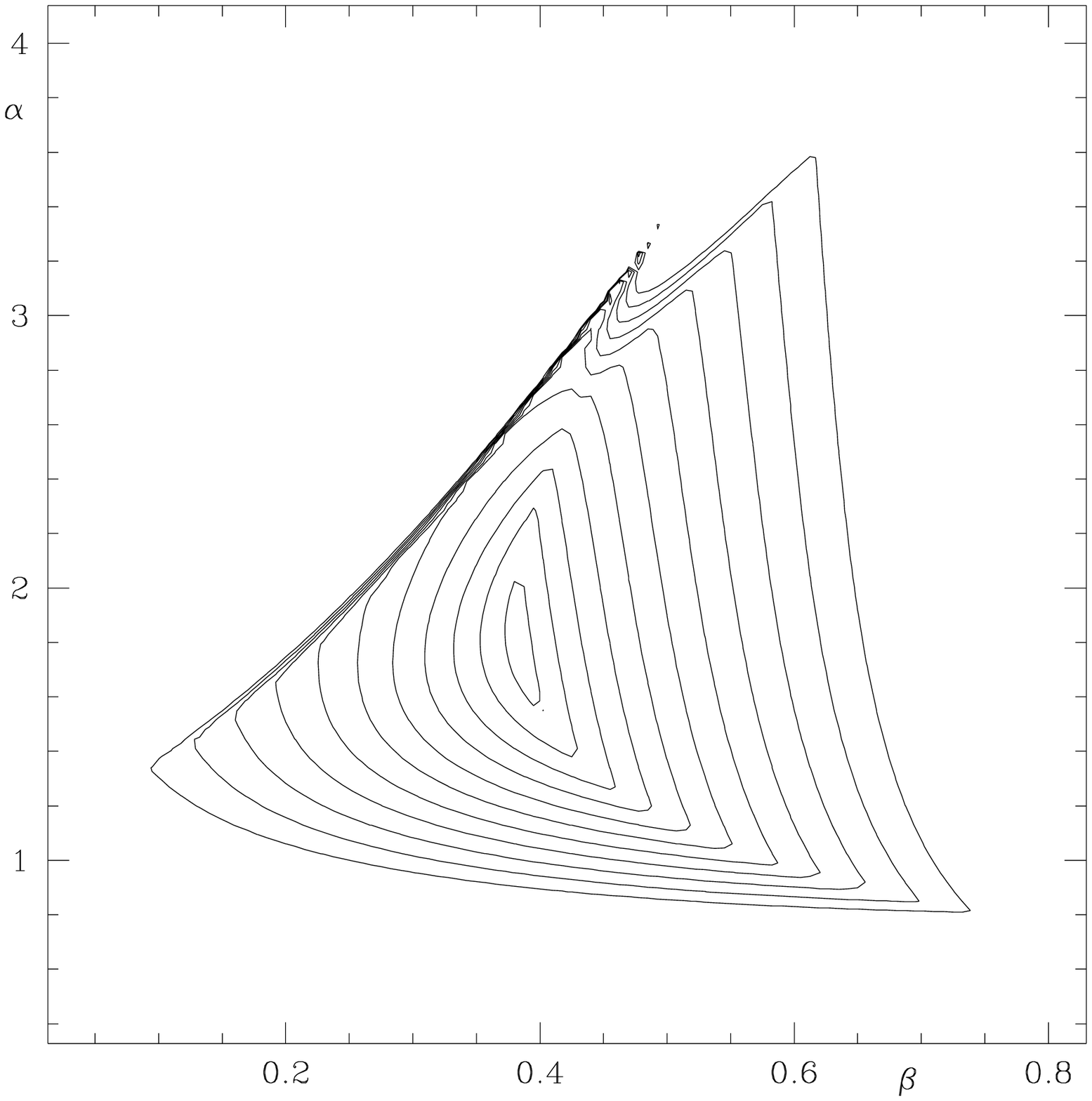}
\caption{}
\end{figure}

\begin{center}
\begin{figure}
\epsfxsize=6in
\epsffile{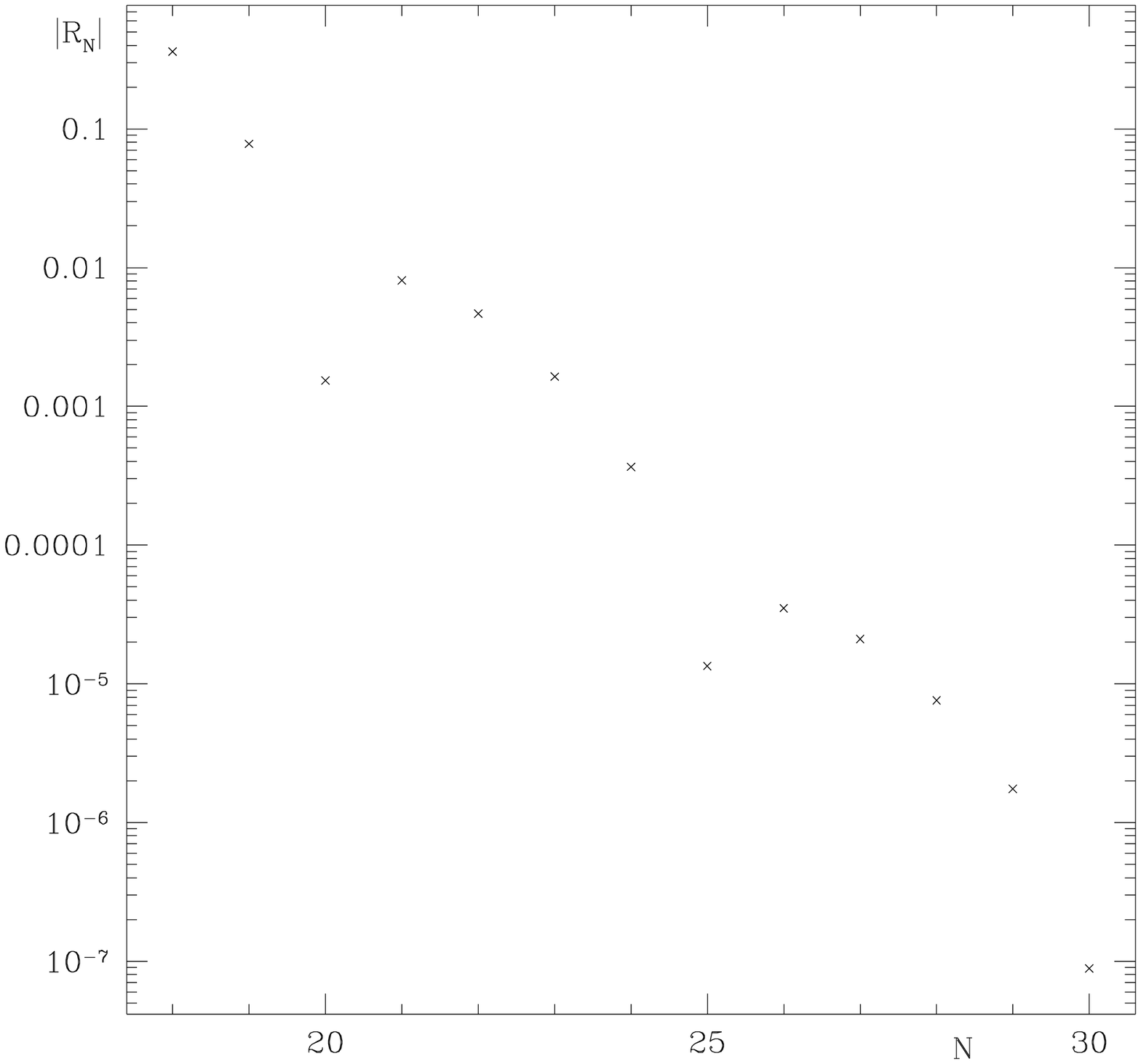}
\caption{}
\end{figure}
\end{center}

\hoffset=-0.5in
\begin{figure}
\begin{center}
\epsfxsize=6in
\epsffile{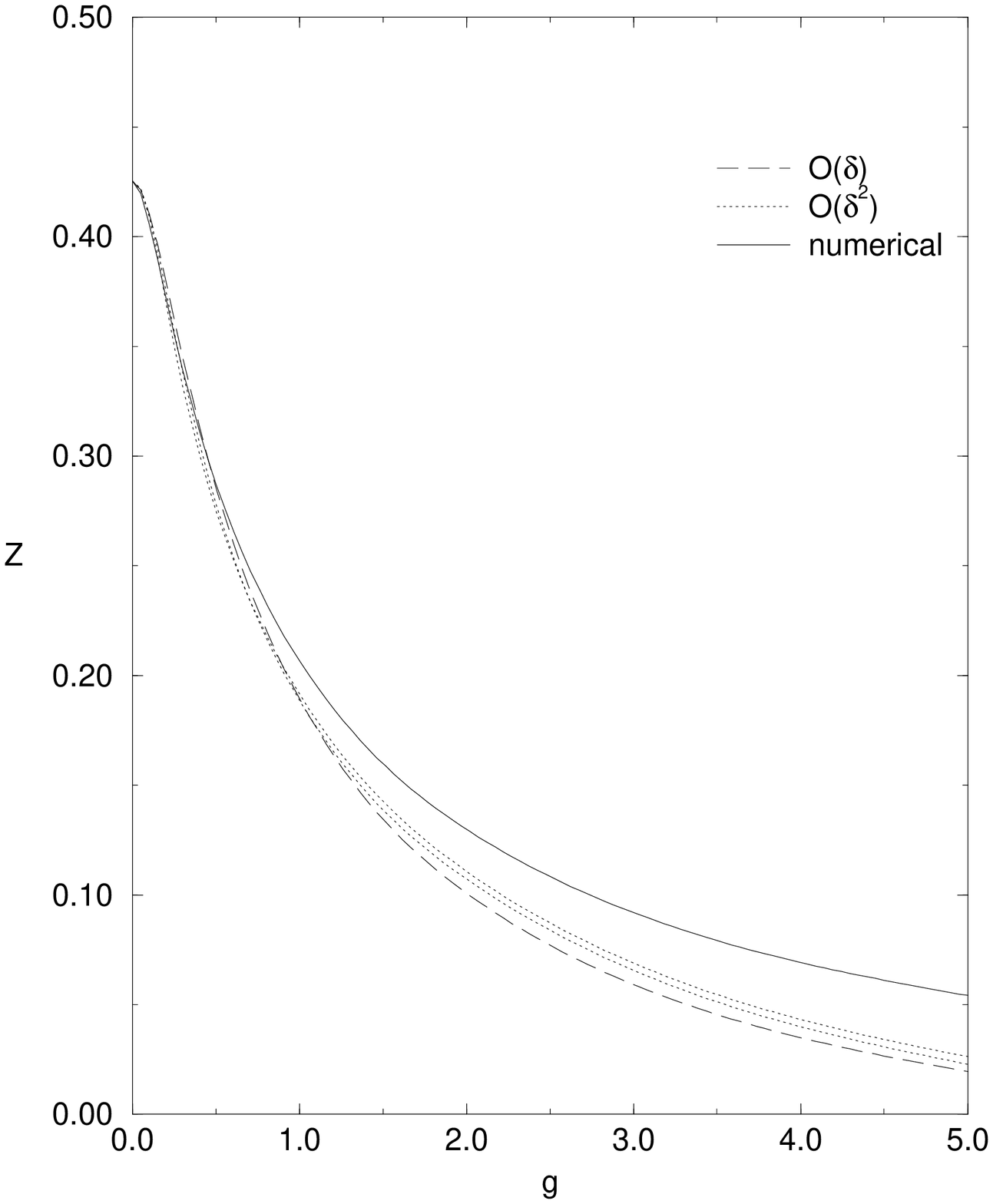}
\vspace{1cm}\caption{}
\end{center}
\end{figure}

\begin{center}
\begin{figure}
\epsfxsize=6in
\epsffile{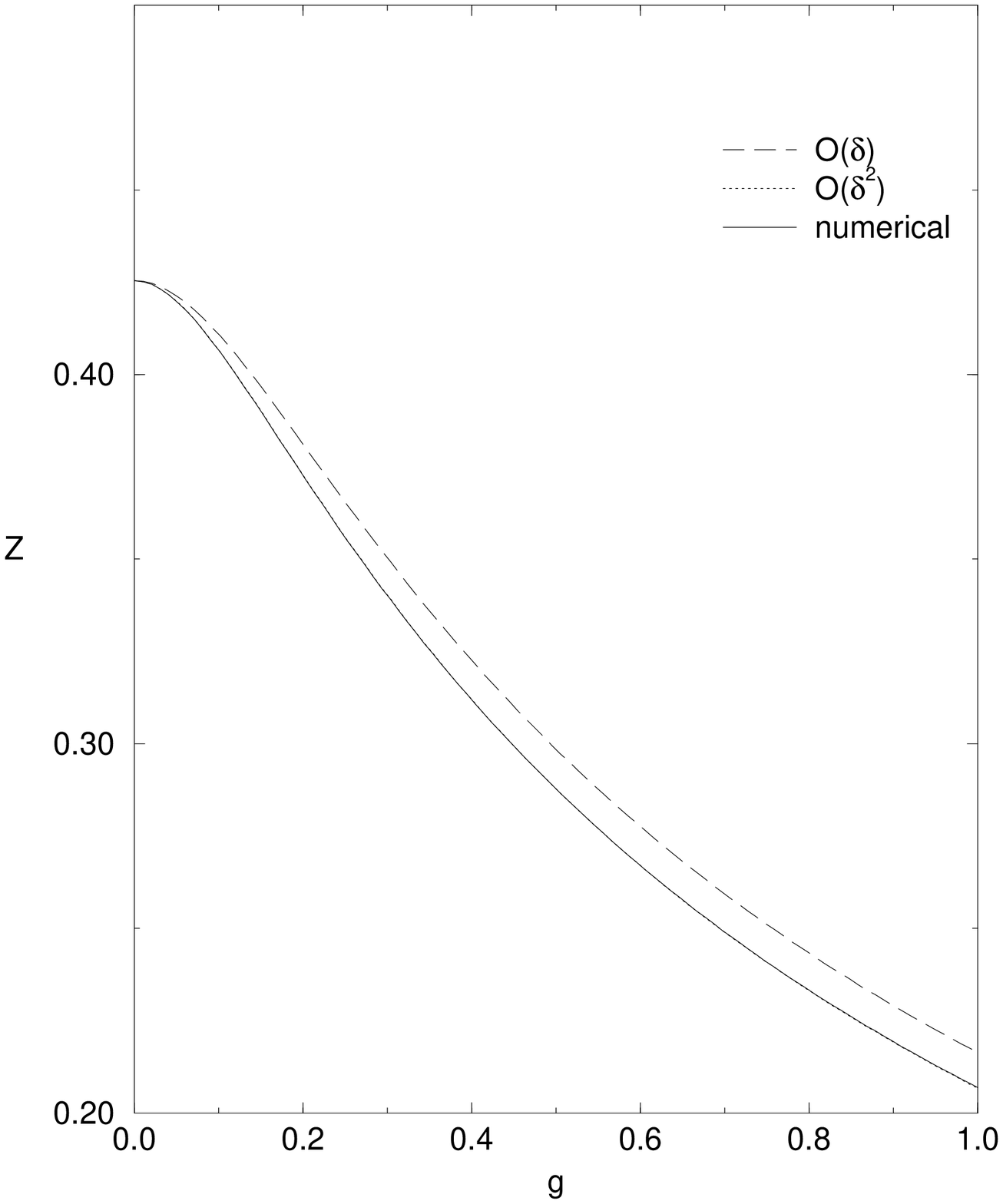}
\vspace{0.5cm}\caption{}
\end{figure}
\end{center}

\begin{center}
\begin{figure}
\epsfxsize=6in
\epsffile{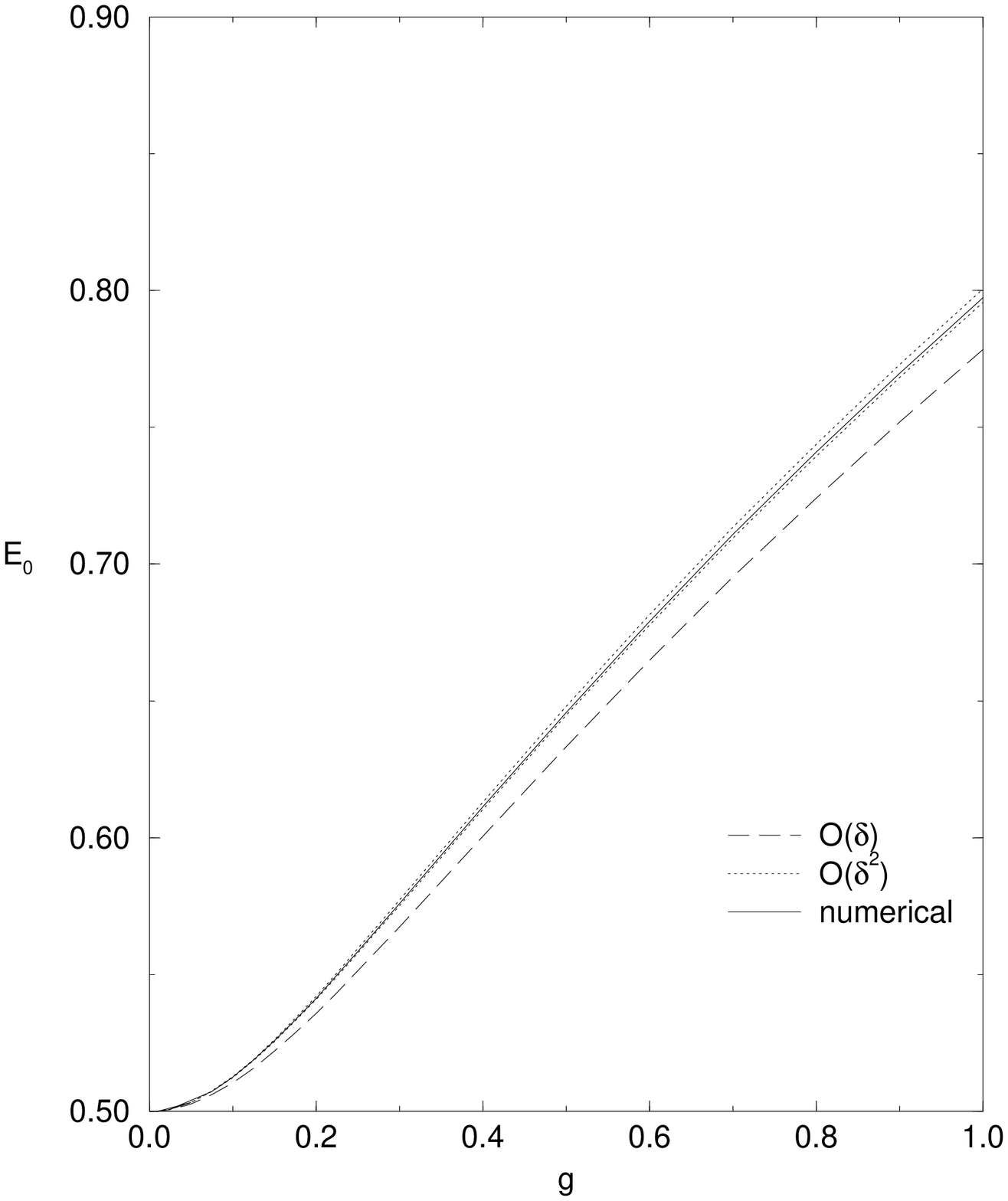}
\vspace{0.5cm}\caption{}
\end{figure}
\end{center}

\end{document}